\documentclass[
prc,%
10pt,%
final,%
notitlepage,%
oneside,%
twocolumn,%
nobibnotes,%
nofootinbib,
superscriptaddress,%
floatfix,%
floatfix,%
showkeys,%
showpacs]%
{revtex4}
\usepackage{color}
\usepackage{amsfonts}
\usepackage{amsbsy}
\usepackage{mathrsfs}
\usepackage{graphicx}
\def\lsim{\mathrel{\rlap{
\lower4pt\hbox{\hskip-3pt$\sim$}}
    \raise1pt\hbox{$<$}}}     
\def\gsim{\mathrel{\rlap{
\lower4pt\hbox{\hskip-3pt$\sim$}}
    \raise1pt\hbox{$>$}}}     
\def\scr#1{\mbox{\scriptsize #1}}
\begin{document}
\title{	
Correlation between global polarization, angular momentum, and flow  
in heavy-ion collisions
} 
\author{Yu. B. Ivanov}\thanks{e-mail: yivanov@theor.jinr.ru}
\affiliation{Bogoliubov Laboratory for Theoretical Physics, 
Joint Institute for Nuclear Research, Dubna 141980, Russia}
\affiliation{National Research Nuclear University "MEPhI", 
Moscow 115409, Russia}
\affiliation{National Research Centre "Kurchatov Institute",  Moscow 123182, Russia} 
\author{A. A. Soldatov}
\affiliation{National Research Nuclear University "MEPhI",
Moscow 115409, Russia}
\begin{abstract}
Possible correlations of the global polarization of $\Lambda$ hyperons 
with  angular momentum and transverse flow in the central region
of colliding nuclei are studied based on a refined estimate of the global polarization. 
Simulations of Au+Au collisions at collision energies $\sqrt{s_{NN}}=$ 6--40 GeV
are performed within the model of the three-fluid dynamics. 
Within the crossover and first-order-phase-transition scenarios 
this refined estimate quite satisfactorily reproduces the experimental STAR data. 
Hadronic scenario fails at high collision energies, $\sqrt{s_{NN}}>$ 10 GeV,
and even predicts opposite sign of the global polarization. 
It is found that the global polarization correlates with neither the angular momentum 
accumulated in the central region nor with directed and elliptic flow. 
At the same time we observed correlation between the angular momentum and 
directed flow in both their time and collision-energy dependence. 
These results suggest that, although initially the angular momentum is the driving force for the vortex generation, later the angular momentum and vortex motion become decorrelated in the midrapidity region. Then the midrapidity angular momentum 
is determined by the pattern of the directed flow and even becomes negative when 
the antiflow occurs. 
At the freeze-out stage, the dominant part of the participant angular momentum 
is accumulated in the fragmentation regions.
\end{abstract}
\maketitle

\section{Introduction}

Experimental observation of the global hyperon polarization in heavy-ion collisions by the STAR
Collaboration at the Relativistic Heavy Ion Collider (RHIC) 
\cite{STAR:2017ckg,Adam:2018ivw,Niida:2018hfw} 
gave us the evidence for the creation of the most vortical fluid ever observed. 
Theoretical simulations within the hydrodynamic approaches 
\cite{Karpenko:2016jyx,Xie:2017upb,Ivanov:2019ern,Ivanov:2019wzg}
and transport models 
 \cite{Li:2017slc,Wei:2018zfb,Shi:2017wpk,Kolomeitsev:2018svb,Vitiuk:2019rfv} 
based on thermal equilibration of the spin degrees of freedom 
\cite{Becattini:2013fla,Becattini:2016gvu,Fang:2016vpj}
succeeded to 
describe the measured global hyperon polarization  
\cite{STAR:2017ckg,Adam:2018ivw,Niida:2018hfw}.
An alternative approach based on the axial vortical effect (AVE)
\cite{Vilenkin:1980zv,Gao:2012ix,Sorin:2016smp} also reasonably reproduced the observed global polarization \cite{Baznat:2017jfj,Sun:2017xhx}. 
Although problems still persist,  see recent review in Ref. \cite{Becattini:2020ngo},
this gives us confidence that our current understanding of the 
heavy-ion dynamics and, in particular, the vortical motion is 
basically compatible with the observed polarization.

This phenomenon of the global polarization is usually related to the Barnett effect 
\cite{Barnett:1915}, i.e. magnetization by rotation, where a fraction of the 
orbital angular momentum associated with the body rotation is transformed into the spin angular momentum. In the Barnett effect the magnetization, i.e. spin alignment, is proportional 
to the angular momentum. On the contrary, the global hyperon polarization 
decreases with collision energy rise, i.e., with an increase of the total angular momentum 
\cite{STAR:2017ckg,Adam:2018ivw,Niida:2018hfw}. This mismatch was explained by  
that the global polarization is measured in central region of the colliding system 
(near the midrapidity), while the angular momentum is mostly accumulated in 
peripheral regions at the freeze-out stage 
\cite{Ivanov:2019ern,Ivanov:2019wzg,Ivanov:2017dff,Ivanov:2018eej}. 
Then the question arises: whether the global polarization in central region
correlates with the angular momentum accumulated in this region? 
In the present paper we study this question.

We start with a more accurate estimate of the global polarization than that
made in the previous paper \cite{Ivanov:2019ern}. Then we compare 
collision-energy dependence and time evolution of 
the global polarization and the angular momentum accumulated in the central region. 
We also compare the above quantities with those of directed and elliptic 
flow to test their possible correlation.  
%
The simulations are performed within the model of the three-fluid dynamics (3FD) \cite{3FD}  
in the energy range of 
the Nuclotron based Ion Collider fAcility (NICA) in Dubna and 
the Beam Energy Scan (BES) program at RHIC.

\section{Polarization in 3FD Model}
\label{Model}

High-energy heavy-ion collisions are characterized by a finite stopping power 
resulting in a counterstreaming regime of baryon-rich matter  
at an early stage of the collision.   
Within the 3FD \cite{3FD} this nonequilibrium regime 
is modeled by two interpenetrating baryon-rich fluids 
initially associated with constituent nucleons of the projectile
(p) and target (t) nuclei. Newly produced particles, predominantly
populating the midrapidity region, are attributed to a fireball (f) fluid.
Each of these fluids is governed by conventional hydrodynamic equations 
coupled by friction terms in the right-hand sides of the Euler equations. 
The physical input of the present 3FD calculations is described in
Ref.~\cite{Ivanov:2013wha}. 
Three different equations of state (EoS's) were used in simulations.  
These are a purely hadronic EoS \cite{gasEOS}  
and two versions of the EoS with the  deconfinement
 transition \cite{Toneev06}, i.e. a first-order phase transition (1PT)   
and a crossover one.

The global polarization of hyperons was measured in the midrapidity region, i.e. at
pseudorapidity $|\eta|<1$ \cite{STAR:2017ckg,Adam:2018ivw,Niida:2018hfw}. 
Similarly to that in Ref. \cite{Ivanov:2019ern}, 
we associate the global midrapidity polarization with the polarization of 
$\Lambda$ hyperons emitted from a central slab of the Au+Au colliding system  
   \begin{eqnarray}
   \label{slab-P_Lambda}
   P_{\Lambda}  
	\simeq 
 \frac{\langle \varpi_{zx} \rangle}{2}
 \left(1 +  \frac{2}{3} 
 \frac{\langle E_{\Lambda}\rangle - m_\Lambda}{m_{\Lambda}} \right),  
   \end{eqnarray}
where $m_\Lambda$ is the mass of $\Lambda$ hyperon, $\langle E_{\Lambda}\rangle$
is energy of the $\Lambda$ hyperon averaged over the central slab, and 
$\langle \varpi_{zx} \rangle$ is the $zx$ component of the thermal vorticity 
averaged over the central slab with the weight of the energy density $\varepsilon$
   \begin{eqnarray}
   \label{en.av.therm.B-vort-T}
  \langle \varpi_{\mu\nu}\rangle (t)  
  &=& \int_{\rm{slab}} 
  d^3 x \;
   [\varpi_{\mu\nu}^{\rm B}({\bf x},t)\;\varepsilon_{\rm B} ({\bf x},t) 
   \cr
   &+&\vphantom{\int dV}\varpi_{\mu\nu}^{\rm f}({\bf x},t)\;\varepsilon_{\rm f}({\bf x},t)]
 \Big/ \langle \varepsilon\rangle (t) .
   \end{eqnarray}
where
   \begin{eqnarray}
   \label{therm.vort.}
   \varpi_{\mu\nu} = \frac{1}{2}
   (\partial_{\nu} \hat{\beta}_{\mu} - \partial_{\mu} \hat{\beta}_{\nu}), 
   \end{eqnarray}
$\hat{\beta}_{\mu}=\hbar\beta_{\mu}$,  $\beta_{\mu}=u_{\nu}/T$, 
$u_{\mu}$ is local four-velocity of a fluid,  and
$T$ is local temperature.  
Here B and f label quantities related to unified baryonic (p and t) fluid and 
the f-fluid, respectively, and 
\begin{eqnarray}
\label{eps-tot-appr}
\varepsilon \simeq  \varepsilon_{\scr B} + \varepsilon_{\scr f}. 
\end{eqnarray}
In Eq. (\ref{slab-P_Lambda}) all quantities are taken at the freeze-out instant.  
Expression (\ref{eps-tot-appr}) is a good approximation 
because of unification of the
baryon-rich fluids and small  relative (between baryon-rich  and fireball fluids)
velocities at the later stages of the collision \cite{Ivanov:2018rrb}.

The above equations need certain comments. 
Eq. (\ref{slab-P_Lambda}) (without averaging over the slab) was derived for polarization  vector averaged over the momentum direction of emitted hyperons \cite{Kolomeitsev:2018svb}. 
In Ref. \cite{Ivanov:2019ern}, where a narrow central slab  was used, 
we neglected the longitudinal motion of the $\Lambda$ hyperon  
in that  slab and therefore approximated $\langle E_{\Lambda}\rangle$
by the mean midrapidity transverse mass, $\langle m_T^\Lambda \rangle_{\scr{midrap.}}$.
As we consider a wider slab in the present calculation (see discussion below), we compute  
$\langle E_{\Lambda}\rangle$ with an account of the longitudinal motion. 
We consider the proper-energy-density weighted vorticity (\ref{slab-P_Lambda})
which allows us to suppress contributions of regions of low-density matter. 
It is appropriate because abundant production of hyperons takes
place in highly excited regions of the system.

A simplified version of the the freeze-out was used in Ref. \cite{Ivanov:2019ern}. 
The freeze-out instant was associated with time, when the energy density 
$\langle \varepsilon (t)\rangle$ averaged over the central slab reached the value of 
freeze-out energy density $\epsilon_{\scr{frz}}= 0.4$ GeV/fm$^3$. 
This parameter is the same for all EoS's and all collision energies. 

In actual calculations of observables a differential, i.e.
cell-by-cell, freeze-out is implemented in the 3FD \cite{Russkikh:2006aa}. 
The freeze-out procedure starts when the local energy density drops down 
to the freeze-out value $\epsilon_{\scr{frz}}$.  
The freeze-out criterion is checked in the analyzed cell and in eight cells 
surrounding this cell. If the freeze-out criterion is met in all cells and
if the analyzed cell is adjacent to the vacuum 
(i.e if at least one of the surrounding cells is ``empty''\footnote{Frozen-out 
cells are removed from the hydrodynamical evolution.}), 
then this considered cell is counted as frozen out. The latter condition 
prevents formation of bubbles of frozen-out matter inside the
dense matter still hydrodynamically evolving. This results in the actual energy density
of frozen-out cell $\varepsilon_{\scr{frz}}$ being lower than $\epsilon_{\scr{frz}}$. 
Thus, $\epsilon_{\scr{frz}}$ has a meaning of a ``trigger'' 
that indicates possibility of the freeze-out.
The physical pattern behind this freeze-out resembles the
process of expansion of a compressed and heated classical fluid into vacuum, 
mechanisms of which were studied
both experimentally and theoretically, see discussion in Ref. \cite{Russkikh:2006aa}. 
The freeze-out is associated with evaporation from the surface of the
expanding fluid. 

\begin{figure}[htb]
\includegraphics[width=6.6cm]{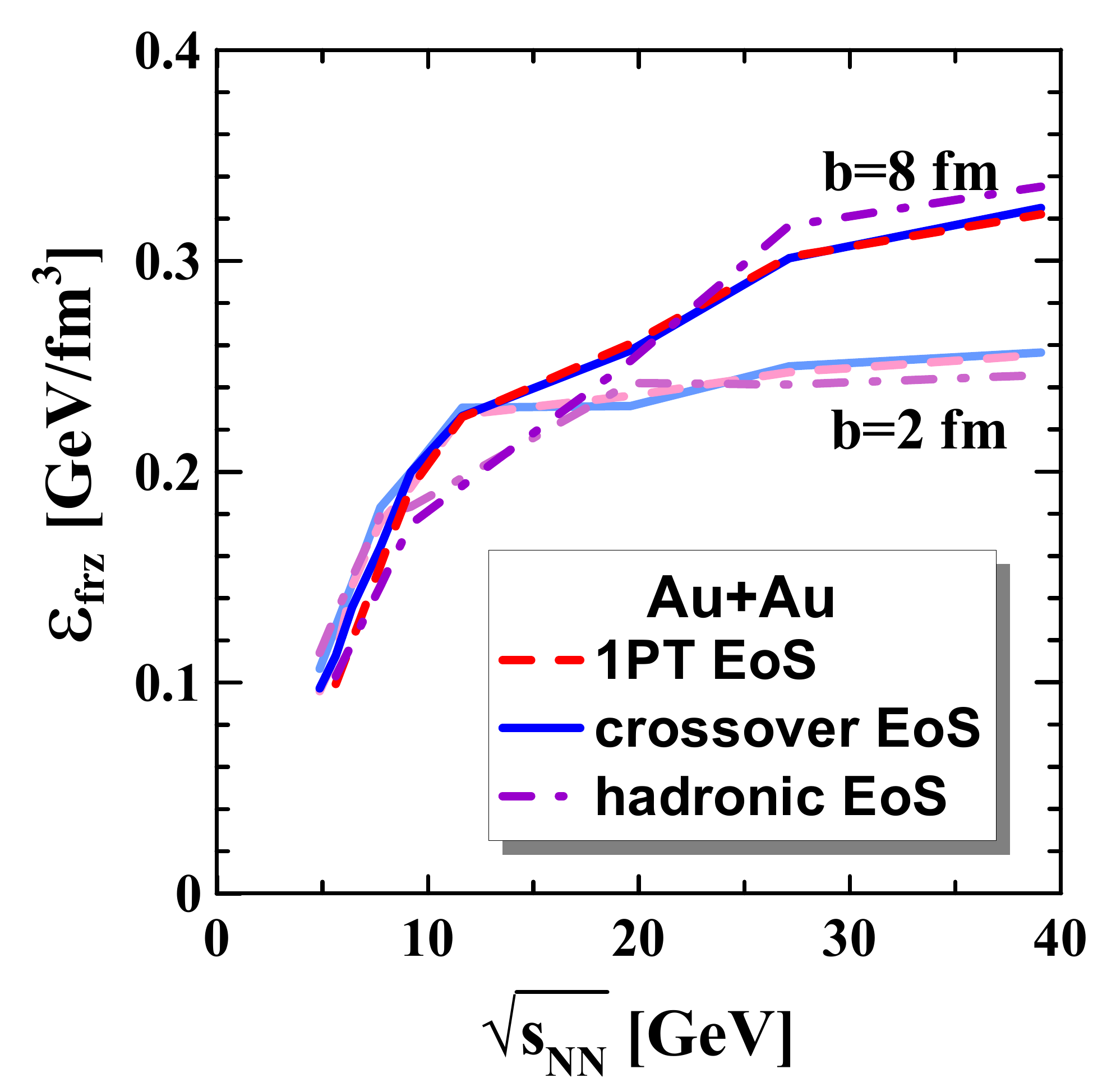}
 \caption{(Color online)
Average actual freeze-out energy density versus collision energy $\sqrt{s_{NN}}$
in Au+Au collisions at impact parameters $b=$ 2 and 8 fm 
calculated with different EoS's. Pale colors are used for $b=$ 2 fm.  
}
\label{fig1}
\end{figure}

The actual value $\varepsilon_{\scr{frz}}$ depends on dynamics of expansion and 
consequently on the collision energy, EoS and impact parameter ($b$). 
This actual freeze-out energy density, averaged over frozen out system, 
is illustrated in Fig. \ref{fig1}.

We performed 3FD simulations of Au+Au collisions at fixed
impact parameters $b=$ 8 fm. This $b$ was taken to roughly 
comply with the STAR centrality selection of 20-50\% \cite{STAR:2017ckg}. 
Glauber simulations of Ref. \cite{Abelev:2008ab} were used to relate 
the experimental centrality and the mean impact parameter. 
In the present calculation we also apply the global freeze-out in the central slab 
but at the actual freeze-out energy density as it is displayed in Fig. \ref{fig1}. 
Detailed discussion of $\varepsilon_{\scr{frz}}$ dependence versus collision 
energy is presented in Ref. \cite{Ivanov:2013yla}. In particular, it is responsible 
for the observed step-like behavior of mean transverse masses as function of 
the collision energy \cite{Ivanov:2013yla}.

\begin{figure}[htb]
\includegraphics[width=8.7cm]{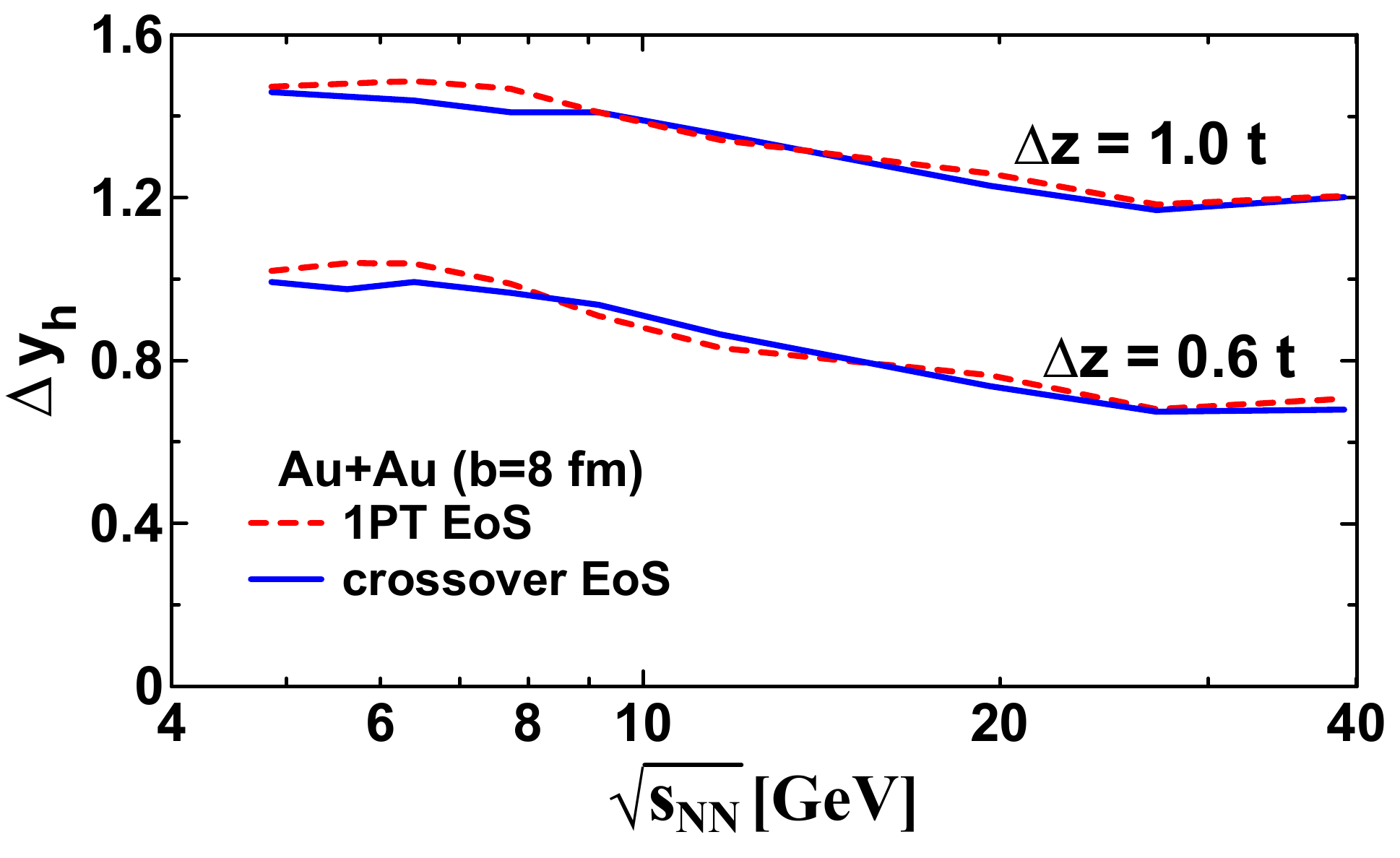}
 \caption{(Color online)
Rapidity width $\Delta y_h$ ($|y_h|< \Delta y_h/2$) of the central slab  
at the freeze-out instant versus collision energy $\sqrt{s_{NN}}$
in Au+Au collisions at impact parameters $b=$ 8 fm 
calculated with various EoS's for two different  
prescriptions of the spatial slab width, Eqs. (\ref{dz05}) and (\ref{dz03}). 
}
\label{fig2}
\end{figure}

Width of the central slab is chosen to model the experimental condition 
$|\eta|<1$. In terms of rapidity of $\Lambda$ hyperons this approximately 
corresponds to $|y|<0.7$. In this estimate we used mean transverse masses 
of $\Lambda$'s measured in Ref. \cite{Adam:2019koz}. We calculate the rapidity 
based on hydrodynamical 4-velocity $u^\mu$ 
   \begin{eqnarray}
   \label{y}
y_h(z,t) = \frac{1}{2} \ln 
\frac{\left\langle u^0+u^3\right\rangle}{\left\langle u^0-u^3\right\rangle} , 
   \end{eqnarray}
where 
   \begin{eqnarray}
   \label{u}
\langle u^\mu\rangle (z,t) =  
\int
  dx^1 dx^2
   [u^\mu_{\rm B}\;\varepsilon_{\rm B}  
   + u^\mu_{\rm f}\;\varepsilon_{\rm f}]
 \Big/ \langle \varepsilon\rangle (z,t) 
   \end{eqnarray}
is the hydrodynamical 4-velocity averaged over $xy$ plane with the weight of 
the proper energy density, cf. Eq. (\ref{en.av.therm.B-vort-T}). 
We use subscript $h$ to indicate that this is a hydrodynamical rapidity 
rather than a true one. 
We define the rapidity width of the central slab as follows 
   \begin{eqnarray}
   \label{dy}
\Delta y_h (t) = y_h(z_{\rm right},t) - y_h(z_{\rm left},t),    
   \end{eqnarray}
where $z_{\rm right}$ and $z_{\rm left}$ are the right and left borders of the 
slab, respectively. In order to approximately keep $\Delta y_h (t) \approx$ 1.4, 
i.e. $|y_h|< \Delta y/2 \approx$ 0.7, we take an 
expanding with time central slab
   \begin{eqnarray}
   \label{dz05}
z_{\rm right}(t) = -z_{\rm left}(t) = 0.5t.     
   \end{eqnarray}
We also did calculations with the central slab
expanding with time as
   \begin{eqnarray}
   \label{dz03}
z_{\rm right}(t) = -z_{\rm left}(t) = 0.3t     
   \end{eqnarray}
in order to simulate the standard STAR selection of the 
midrapidity region $|y| <$ 0.5, i.e. $\Delta y_h (t) \approx$ 1. 
Results of estimations of $\Delta y_h$ at the freeze-out instant
according to Eqs. (\ref{y})--(\ref{dz03}) are shown in Fig. \ref{fig2}. 
As seen, the results are not perfect because $\Delta y$ depends on 
$\sqrt{s_{NN}}$, but it approximately stays near desired values. 
%

%
\begin{figure}[bht]
\includegraphics[width=8.9cm]{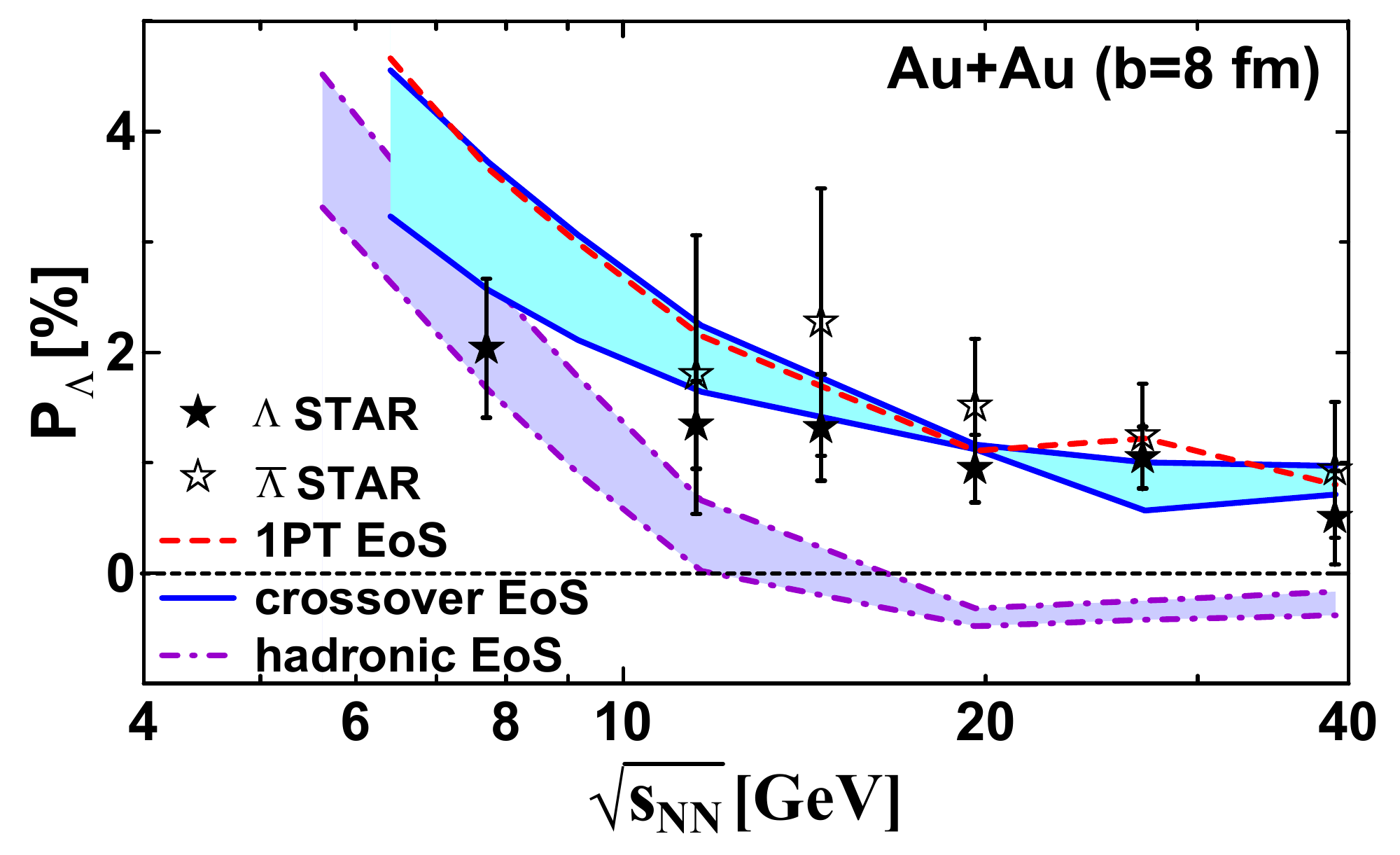}
 \caption{(Color online)
 Global 
polarization of $\Lambda$ hyperons in Au+Au collisions at $b=$ 8 fm as 
function of collision energy $\sqrt{s_{NN}}$.  
Shaded bands for the crossover EoS and hadronic EoS 
indicate polarization sensitivity to 
choice of the central slab: the upper borders of these bands correspond to choice (\ref{dz05}) 
and the lower borders - to choice (\ref{dz03}). 
STAR data on global $\Lambda$ and $\bar{\Lambda}$ polarization \cite{STAR:2017ckg} are also displayed. 
}
\label{fig3}
\end{figure}

The above described improvements of the polarization calculation increase the polarization 
as compared to that reported in Ref. \cite{Ivanov:2019ern}. 
The results of this refined estimate are presented in Fig. \ref{fig3}. 
In order to get an impression of the effect of rapidity window $\Delta y_h$ on the 
resulting polarization, we present calculations with two choices of the central slab: 
Eqs. (\ref{dz05}) and (\ref{dz03}). The polarization increases when the rapidity window 
expands because the polarization is higher at non-central rapidities. 
While the global polarization predicted by the crossover and 1PT EoS's is very 
similar, this is not the case for the hadronic EoS. At high collision energies 
the hadronic-EoS results even in the negative polarization, which looks counter-intuitive 
from the point of view of spin alignment along the angular momentum. This 
will be analyzed in more detail in the next section.

Overall agreement of the present estimate with the STAR data \cite{STAR:2017ckg}
on the $\Lambda$ polarization is quite reasonable. The $\bar{\Lambda}$ polarization 
is very close to the $\Lambda$ one, therefore we do not present it here. 
Note that feed-down contribution to $\Lambda$ polarization due to 
decays of higher mass hyperons is not taken into account in the present estimate. 
This feed-down results in about 10--15\% decrease
of the resulting polarization, as demonstrated in Refs. 
\cite{Karpenko:2016jyx,Becattini:2020ngo,Becattini:2016gvu,Becattini:2019ntv}.

\section{Correlations between polarization, angular momentum and flow}
\label{Correlations}

The method described above is not optimal for calculating the polarization, which can then be compared with experimental data. At the same time it provides a definite advantage 
for study of correlation between the polarization 
and the angular momentum. In this method we can compare the global polarization with the 
angular momentum accumulated in the same space region. 
The angular momentum accumulated in the slab region is defined as 
   \begin{eqnarray}
   \label{J-tot}
J = \int_{\rm slab} d^3 x 
\sum_{\alpha=\rm{p,t,f}} (z\; T^{\alpha}_{10} - x\; T^{\alpha}_{30}),    
   \end{eqnarray}
where $T^{\alpha}_{\mu\nu}$ is the energy-momentum tensor of the $\alpha$(=p,t,f) fluid  
and has the conventional hydrodynamical form, $z$ is the beam axis, 
$(x,z)$ is the reaction plane of the colliding nuclei. 
In view of further comparison 
of the polarization with slope of the directed flow in the center of colliding system, 
we consider a narrower ``$|y_h|\lsim 0.5$'' slab, see Eq. (\ref{dz03}), in this section.
We also limit our further consideration to crossover and 1PT scenarios as the most relevant to the experimental data, see Fig. \ref{fig3}.

\begin{figure}[bht]
\includegraphics[width=8.7cm]{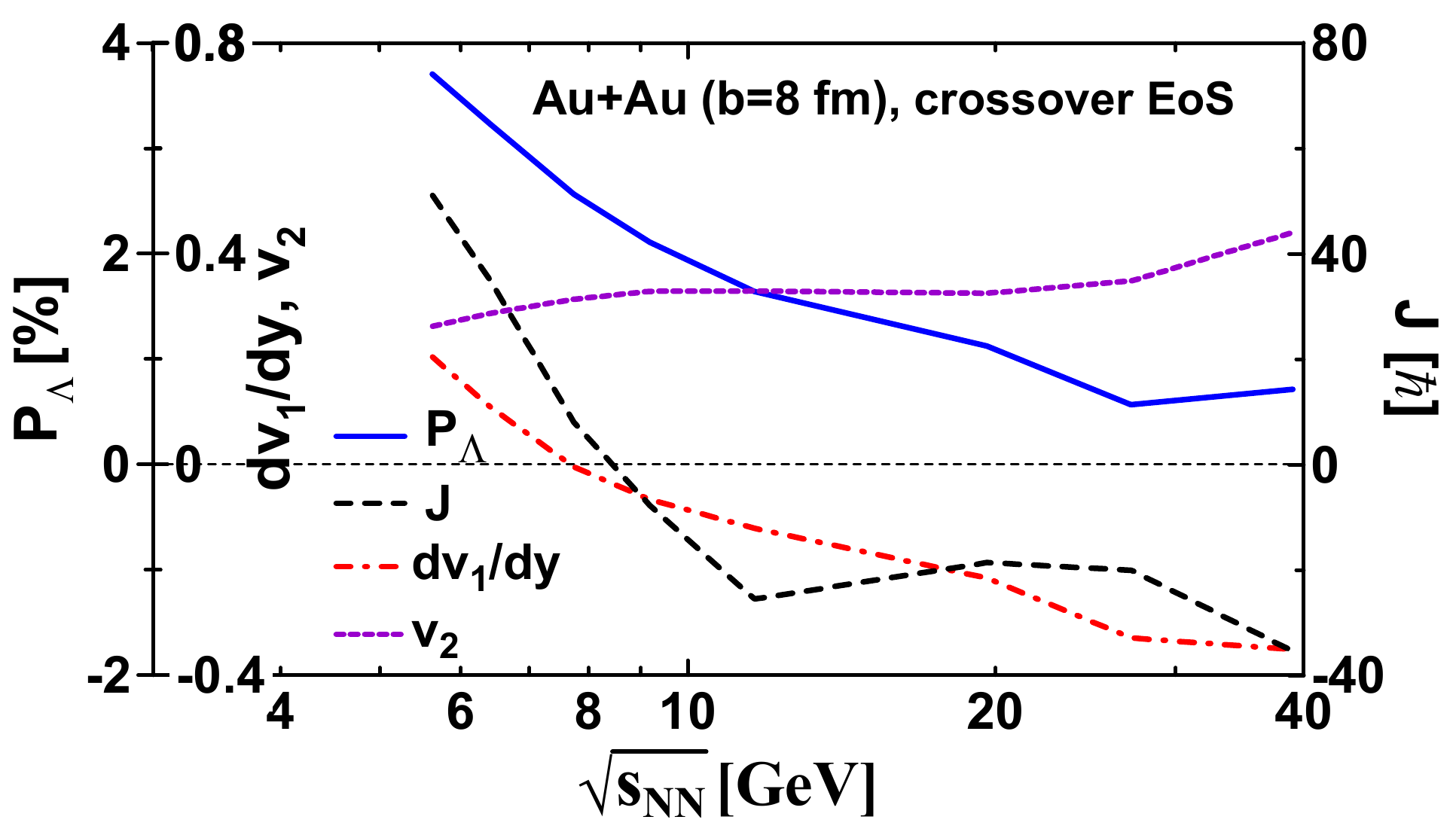}
 \caption{(Color online)
Global polarization of $\Lambda$ hyperons ($P_\Lambda$), angular momentum ($J$) 
accumulated in the ``$|y_h|\lsim 0.5$'' slab, see Eq. (\ref{dz03}),  slope of the 
directed flow of matter ($dv_1/dy$) and elliptic flow ($dv_2$) 
at the freeze-out instant   
in Au+Au collisions at $b=$ 8 fm as functions of the collision energy $\sqrt{s_{NN}}$.  
Calculations are performed with the crossover EoS. 
}
\label{fig4}
\end{figure}

\begin{figure}[bht]
\includegraphics[width=8.7cm]{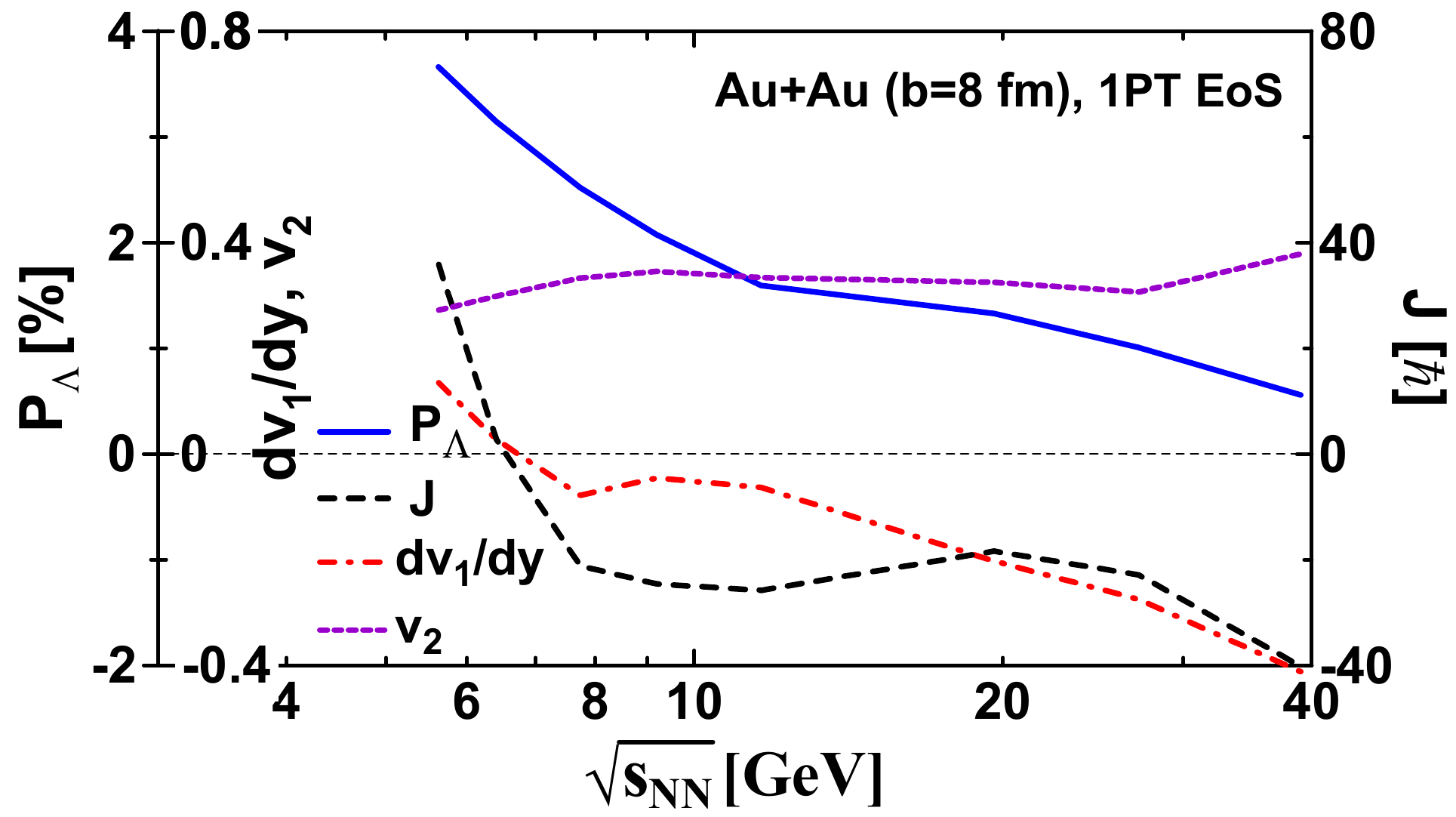}
 \caption{(Color online)
The same as in Fig \ref{fig4} but for the 1PT EoS. 
}
\label{fig5}
\end{figure}

The directed flow of the matter is calculated as 
   \begin{eqnarray}
   \label{v1}
v_1 (y_h,t) = \left\langle \frac{ u^1}{\sqrt{(u^1)^2+(u^2)^2}}\right\rangle   
   \end{eqnarray}
in terms of the hydrodynamical 4-velocities $u^\mu(x)$, $\left\langle ...\right\rangle$ 
means averaging over $xy$ plane with the weight of the proper energy density and 
summation over fluids similarly to Eq. (\ref{u}). 
The corresponding rapidity $y_h$ is defined by Eq. (\ref{y}) in terms of the same 
hydrodynamical 4-velocities. 
The slope of $v_1$ in the center of colliding system, i.e. at ``midrapidity'' $y_h=0$, 
is calculated as 
   \begin{eqnarray}
   \label{dv1/dy}
\frac{d v_1(t)}{dy_h} = 
\frac{v_1(y_{\rm right},t) - v_1(y_{\rm left},t)}{y_{\rm right}(t) - y_{\rm left}(t)}, 
   \end{eqnarray}
where $y_{\rm right}(t)$ and $y_{\rm left}(t)$ are $y_h$ rapidities at 
the right and left borders of the slab, respectively, cf. Eq. (\ref{dy}).
Elliptic flow of the matter at ``midrapidity'' $y_h=0$ is defined similarly
   \begin{eqnarray}
   \label{v2}
v_2(t) = \left\langle \frac{(u^1)^2-(u^2)^2}{(u^1)^2+(u^2)^2}\right\rangle.    
   \end{eqnarray}
Here the $\left\langle ...\right\rangle$ 
averaging is done over $z=0$ plane. 
The above defined quantities at the freeze-out instant are presented in Figs. 
\ref{fig4} (for the crossover EoS) and \ref{fig5} (for the 1PT EoS)
as functions of the collision energy.

%
%
%

Figures \ref{fig4} and \ref{fig5} demonstrate that the global polarization does not 
correlate with the angular momentum ($J$) accumulated in the central slab. 
This angular momentum first decreases with the collision energy rise 
and then flattens at higher energies. 
Moreover, the angular momentum becomes negative at higher collision energies
while the global polarization remains positive. This behavior of the angular momentum 
 in the central slab is completely different from that of the angular momentum 
accumulated in the whole participant region. The latter steadily increases with 
the collision energy rise \cite{Ivanov:2019ern,Ivanov:2019wzg}. 
Absence of correlation between the global polarization and the angular momentum 
was found already in Ref. \cite{Karpenko:2016jyx}. However, there the authors 
considered the angular momentum accumulated in the whole participant region which 
steadily rises with the collision energy. 

In fact, the absence of correlation between the global polarization and 
the angular momentum is not surprising.  
The polarization is intrinsically related to the vorticity while the rotation of the fluid
can be vortex-free. In such vortex-free rotation the vorticity is present only 
in close vicinity of the axis of the rotation, if there is the matter in this vicinity. 
Thus, the angular momentum 
can be arbitrarily large while the global polarization may be generated only in 
the narrow region around the  rotation axis. Moreover, there could be local islands, 
where the matter vortically rotates in the opposite direction to the global 
rotation. Then the angular momentum and the global polarization can have 
opposite signs, as it is the case in the Au+Au collisions at $b=$ 8 fm, 
see Figs. \ref{fig4} and \ref{fig5}. 
Such an island structure in the $xz$ plane was observed in many simulations 
of nuclear collisions 
\cite{Becattini:2015ska,Jiang:2016woz,Li:2017slc,Kolomeitsev:2018svb,Vitiuk:2019rfv,Ivanov:2017dff,Ivanov:2018eej}. 

\begin{figure}[!htb]
\includegraphics[width=8.7cm]{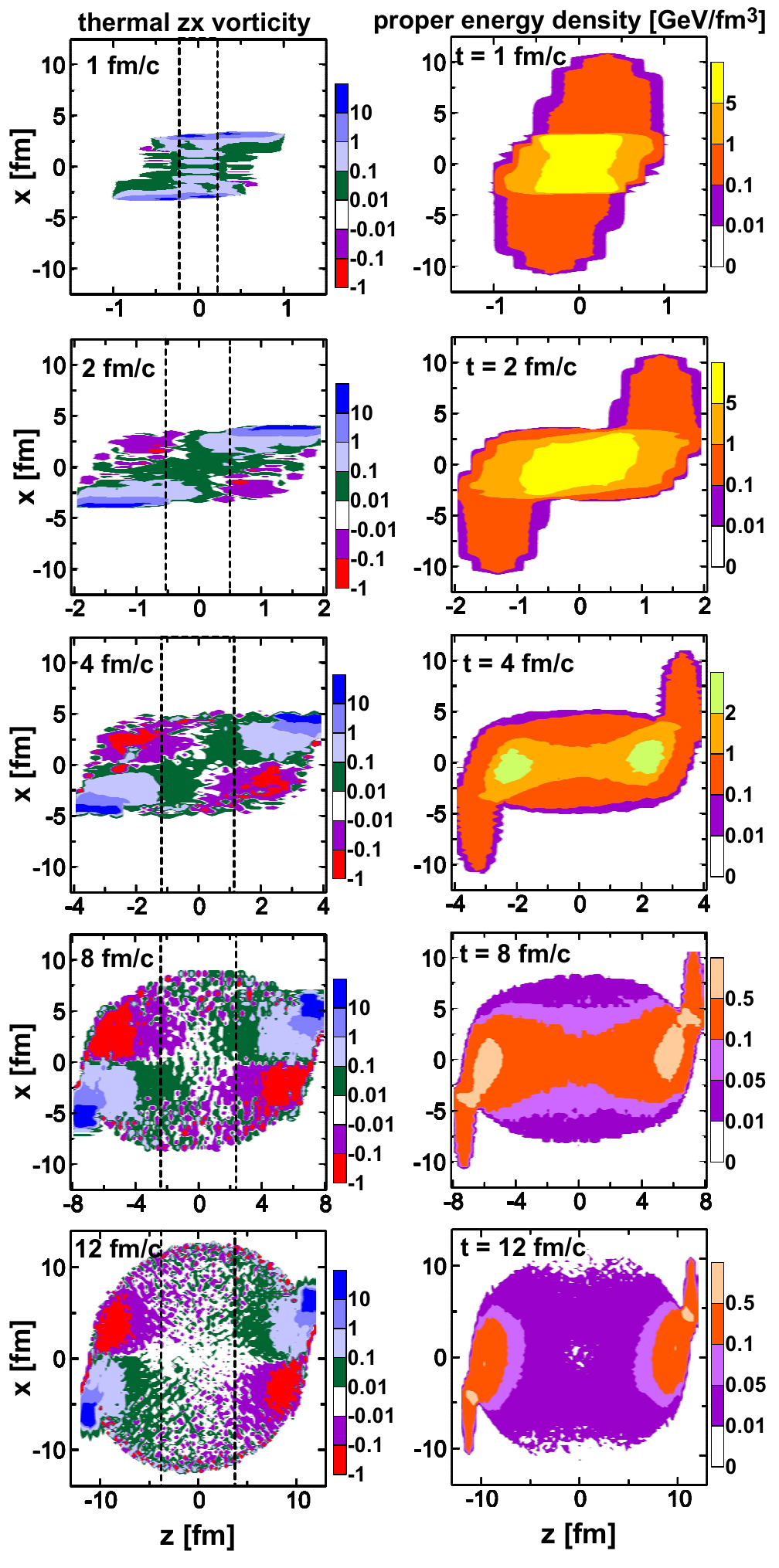}
 \caption{(Color online)
The 
 thermal $zx$ vorticity (left column) and the proper energy density (right column)  
in the reaction plane at various time instants 
in the semi-central ($b=$ 8 fm) Au+Au collision at $\sqrt{s_{NN}}=$ 19.6 GeV. 
Calculations are done with the crossover EoS. $z$ axis is the 
beam direction. Note different scale along the $z$ axis at different time instants. 
Dashed lines around origin of the $x$ axis in the left panels indicate borders of 
the central slab corresponding to the $|y_h|\lsim 0.5$ condition. 
}
\label{fig10}
\end{figure}

This island structure in the $xz$ plane is illustrated in Fig. \ref{fig10}, where 
we present snapshots of the time evolution of the thermal $zx$ vorticity (left column) 
and the proper energy density (right column)  
in the reaction plane 
in the semi-central ($b=$ 8 fm) Au+Au collision at $\sqrt{s_{NN}}=$ 19.6 GeV.
In Figs. \ref{fig4} and \ref{fig5} we analyze the polarization and angular momentum 
accumulated in the central slab corresponding to the $|y_h|\lsim 0.5$ condition.
Borders of this central slab are indicated by 
dashed lines around origin of the $x$ axis in the left panels of Fig. \ref{fig10}.  
Quadrupole structure of vorticity field starts to form already at early stage  
of the collision, i.e. at $t\lsim$ 2 fm/c, but the vortex, rotating along  
the total angular momentum of the system, still dominates in the central slab. 
At later stages the strong positive 
vorticity (i.e. corresponding to the sign of the total angular momentum of the system) 
is pushed out to target and projectile fragmentation regions, which suggests increase of 
the polarization from the midrapidity to the fragmentation regions, as it was argued in  
Refs. \cite{Ivanov:2019ern,Ivanov:2019wzg}. 
While the polarization 
in the central slab is a matter of a delicate 
balance between vortexes and anti-vortexes, i.e. those rotating against the 
total angular momentum.  
The region of high energy density is slightly inclined towards spectators. 
This enhances contribution of the vortexes in the global polarization because the 
hyperons are predominantly produced in highly excited regions. 
The delicate balance between vortexes and anti-vortexes depends on the used EoS. 
Apparently, the hadronic EoS changes this balance in such a way that the polarization 
becomes even slightly negative at high collision energies, see Fig. \ref{fig4}.

The pattern of the kinematic vorticity, i.e. that without extra 
factor $1/T$  in Eq.  (\ref{therm.vort.}), is very similar to that of the thermal vorticity. 
Of course, neither the thermal vorticity nor the kinematic vorticity do not 
characterize the angular momentum, 
in particular, because of possible vortex-free rotation mentioned above. However, 
they give us impression of direction of the rotation in the system. 
As seen from the vorticity field in the central slab, there is no obviously 
preferable direction of the matter rotation. Therefore, 
the angular momentum accumulated in the central slab is also a matter of a delicate 
balance between vortexes, anti-vortexes and possible vortex-free rotation of the matter that 
cannot be seen in Fig. \ref{fig10}. Moreover, $|J|\approx 20 \hbar$ at 
$\sqrt{s_{NN}}=$ 19.6 GeV in the central slab, see Figs. \ref{fig4} and \ref{fig5}, 
whereas the total angular momentum accumulated in the participant region is 
$J_{\rm participants}\approx 2\cdot 10^4 \hbar$, 
see Refs. \cite{Ivanov:2019ern,Ivanov:2019wzg}, i.e. three orders of magnitude higher 
than $|J|$.

\begin{figure}[htb]
\includegraphics[width=8.2cm]{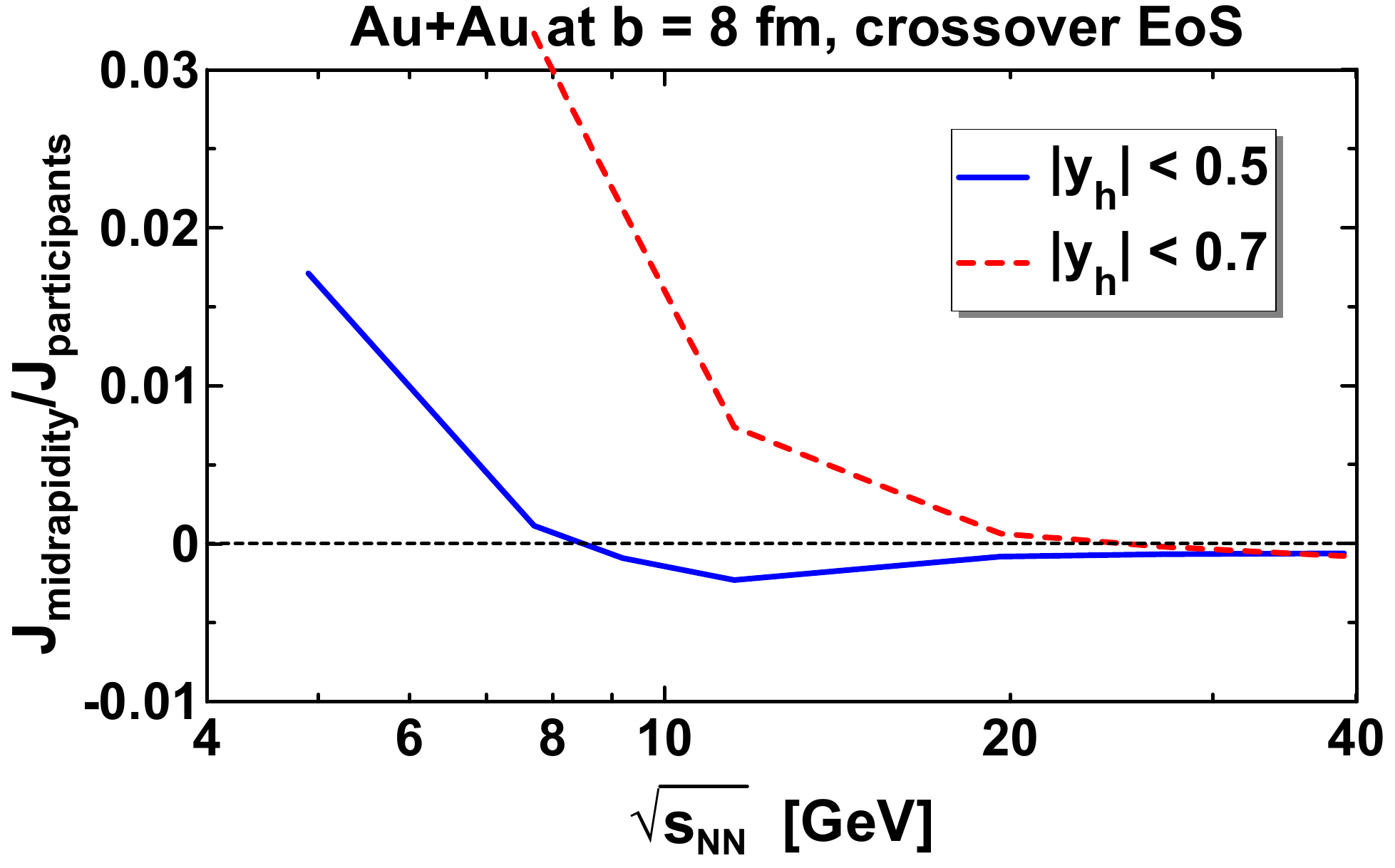}
 \caption{(Color online)
Ratio of the angular momentum in the midrapidity region over   
the angular momentum accumulated by all participants  
at the freeze-out instant 
in semi-central ($b=$ 8 fm) Au+Au collisions as function of $\sqrt{s_{NN}}$. 
Results are presented for two midrapidity regions (central slabs) corresponding 
to the $|y_h|\lsim 0.5$ and $|y_h|\lsim 0.7$ conditions.
Calculations are done with the crossover EoS. 
}
\label{fig11}
\end{figure}

Similar situation takes place at other collision energies, see Fig. \ref{fig11}. 
The value of $|J|$ at the freeze-out instant does not exceed few percent of 
the total participant angular momentum in all considered energy range. 
As seen from Fig. \ref{fig11}, the midrapidity angular momentum  
rapidly decreases with collision energy rise and even becomes negative at high  
collision energies. The midrapidity angular momentum is larger in the wider 
($|y_h|\lsim 0.7$) midrapidity range than that in the narrower ($|y_h|\lsim 0.5$) one.  
This once again indicates that the angular momentum is concentrated in fragmentation regions 
at the freeze-out instant. 
As this midrapidity angular momentum does not correlate with global polarization and 
hence with the vortical motion we can assume that it can be associated with  
collective flow pattern in the midrapidity region.  

Figs. \ref{fig4} and \ref{fig5} also demonstrate collective flow. 
Scales of 
the slopes of the directed flow ($dv_1/dy$) and elliptic flow ($v_2$) in Figs. 
\ref{fig4} and \ref{fig5} noticeably exceed those observed in experiment. 
This is because the  
considered quantities characterize the medium rather than observed particles. 
The flow of observed particles is considerably smeared out by thermal spread 
and resonance decays \cite{Ivanov:2014zqa,Konchakovski:2014gda}. 
The angular momentum characterizes the medium. Therefore, we compare it with 
the flow of the medium rather than specific particles.

As seen, the slope of the directed flow ($dv_1/dy$) does not correlate with 
the polarization but does correlate with the slab angular momentum. The slope 
and the angular momentum even simultaneously change their signs. 
This indicates that tilting the central fireball, which caurses the anti-flow 
\cite{Brachmann:1999xt,Csernai:1999nf}, 
 is accompanied by a change in its angular momentum.  
At the same time the elliptic flow ($v_2$) correlate with neither the polarization nor 
the angular momentum. 

%
\begin{figure}[bht]
\includegraphics[width=8.6cm]{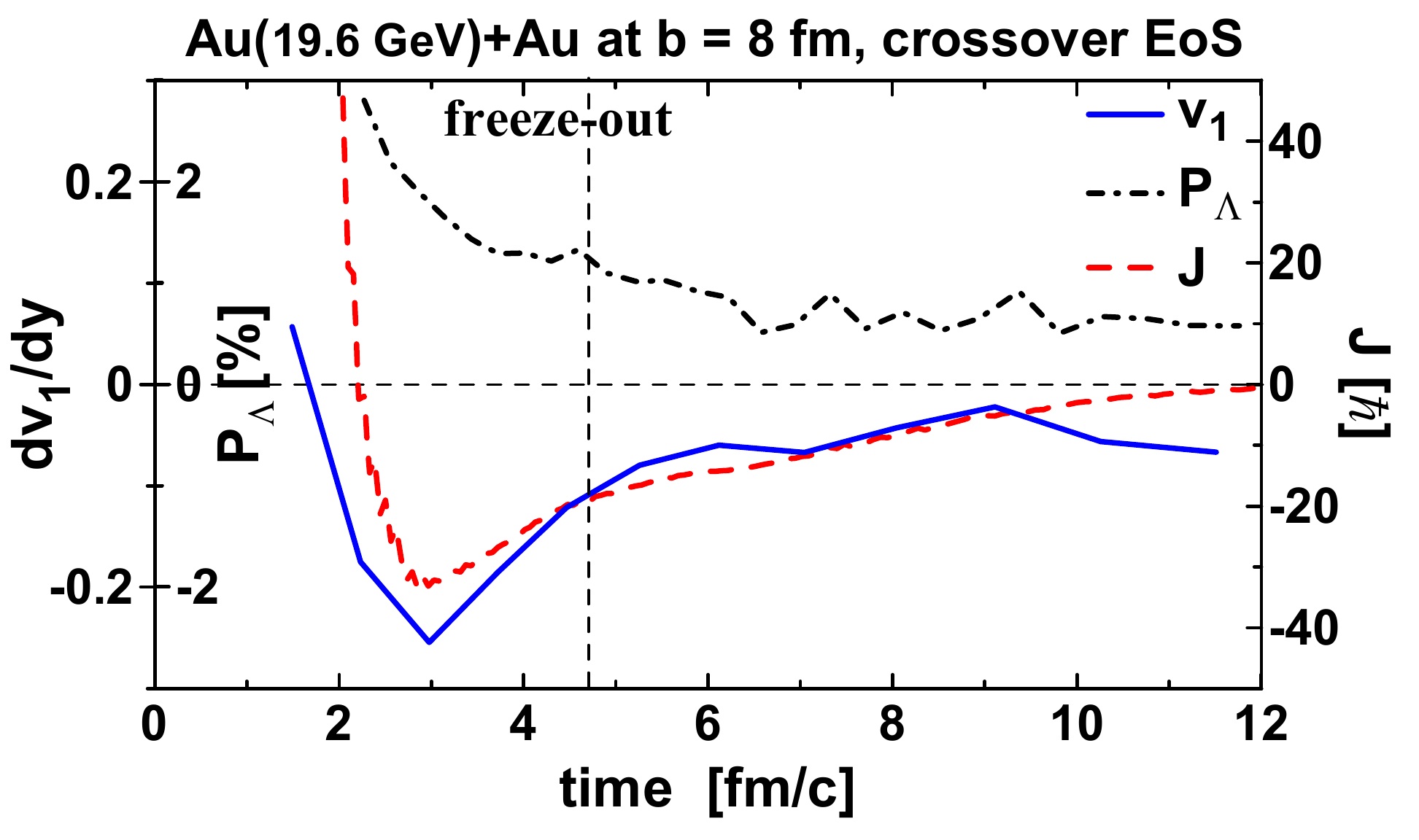}
 \caption{(Color online)
Time evolution of the 
global polarization of $\Lambda$ hyperons ($P_\Lambda$), angular momentum ($J$) 
accumulated in the ``$|y_h|\lsim 0.5$'' slab, see Eq. (\ref{dz03}), and slope of the 
directed flow of matter ($dv_1/dy$) 
in Au+Au collisions at $\sqrt{s_{NN}}=$ 19.6 GeV and $b=$ 8 fm.  
The vertical dashed line indicate the freeze-out instant. 
Calculations are performed with the crossover EoS. 
}
\label{fig7}
\end{figure}
%
%
\begin{figure}[bht]
\includegraphics[width=8.6cm]{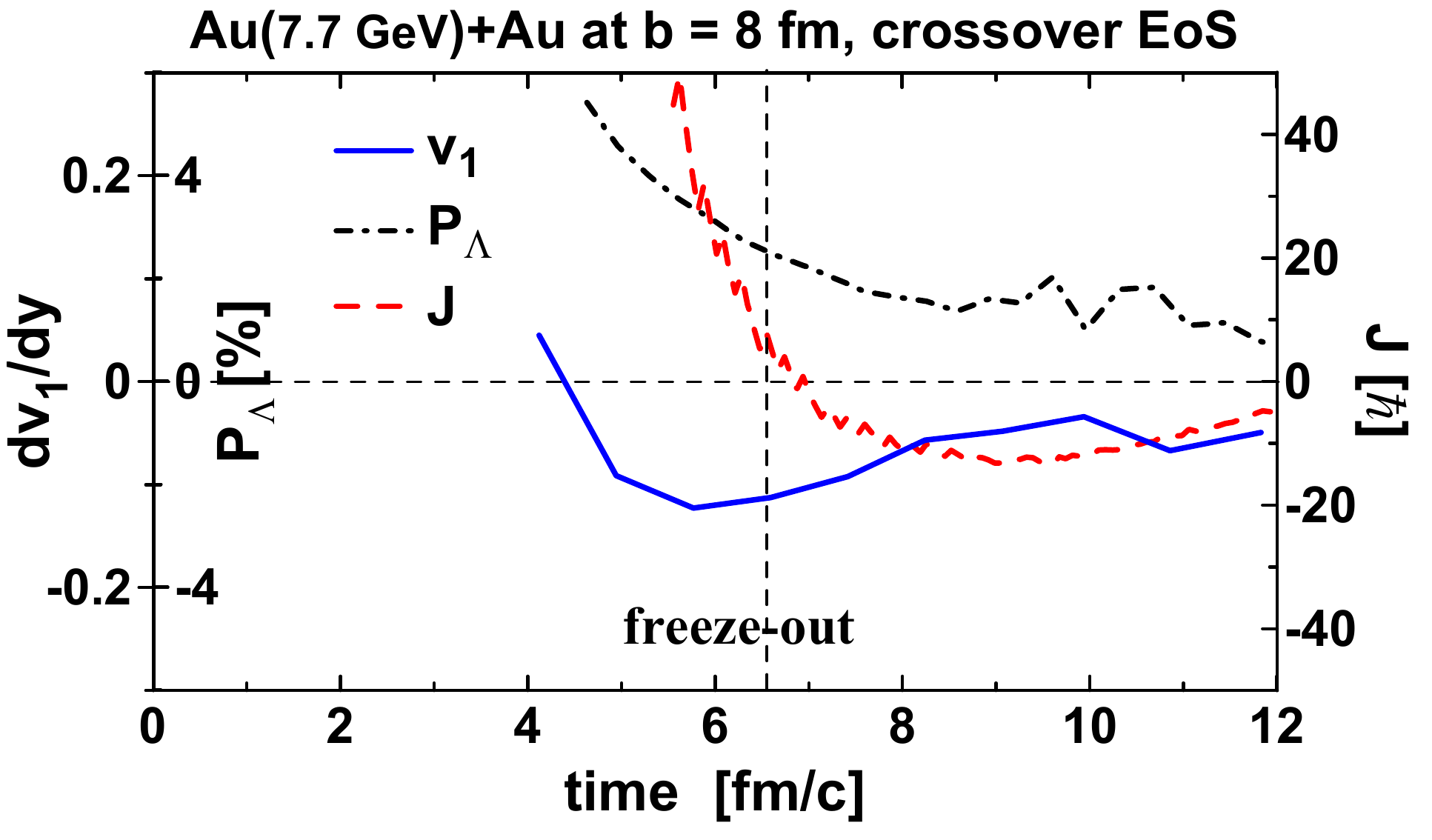}
 \caption{(Color online)
The same as in Fig. \ref{fig7} but for $\sqrt{s_{NN}}=$ 7.7 GeV.  
}
\label{fig8}
\end{figure}

In order to check whether this flow--angular-momentum correlation in their  
$\sqrt{s_{NN}}$ dependence is accidental or not, we also consider their time 
dependence at various energies. Examples of such time dependence are presented 
in Figs. \ref{fig7} and \ref{fig8}.  The time dependence indicates that indeed there is 
a correlation between the $v_1$ slope and angular momentum, which is less  
spectacular at lower collision energies, see Fig. \ref{fig8}. 
Apparently, this is because the chosen width of the central region, 
$|y_h|\lsim 0.5$, is too large at lower collision energies in view of comments below. 
A correlation between $v_1$ flow and polarization is also absent in their 
time dependence.  

Note that the slope of the directed flow ($dv_1/dy$) in Figs. \ref{fig7} and \ref{fig8}
at the freeze-out instant is slightly different from that presented in Fig. \ref{fig4}. 
In Figs. \ref{fig7} and \ref{fig8}, the directed flow is calculated at the Lagrangian 
stage of the code \cite{3FD} in terms of test particles and with smaller step in $y_h$
rapidity than in (\ref{dv1/dy}), 
while that in Fig. \ref{fig4}, see Eq. (\ref{dv1/dy}),    
 is computed on a fixed grid (so called Euler step of the scheme). The Lagrangian 
calculation is more accurate but more time consuming than the Euler one. 
Therefore, we performed this Lagrangian calculation with larger time step than the 
Euler one. To accurately fix the freeze-out instant (as in Fig. \ref{fig4}), we 
need a finer time step.

\section{Summary}
\label{Summary}

Possible correlations of the global polarization of $\Lambda$ hyperons 
with the angular momentum and transverse flow in the central region
of colliding nuclei are studied 
based on refined estimate of the global polarization within the 3FD model. 
In the present approach the global polarization is associated with the 
$\Lambda$ polarization in  the central region of colliding nuclei. 
Within the crossover and first-order-phase-transition scenarios 
this estimate quite satisfactorily reproduces
the experimental STAR data \cite{STAR:2017ckg}, especially its collision-energy dependence.
The purely hadronic scenario fails at high collision energies, $\sqrt{s_{NN}}>$ 10 GeV,
and even predicts the opposite sign of the global polarization.

It is found that the global polarization correlates with neither the angular momentum 
accumulated in the central region nor with directed and elliptic flow. 
Contrary to the polarization, the angular momentum accumulated in the central region
even changes its sign at later stages of nuclear collisions at high collision energies. 
At the same time we detected correlation between the angular momentum and 
directed flow. The midrapidity slope of the directed flow 
and the angular momentum even almost simultaneously change their signs.

The obtained results indicate that, although initially the angular momentum is the driving force for the vortex generation, later the angular momentum and vortex motion become decorrelated in the midrapidity region. Then the midrapidity angular momentum 
is determined by the pattern of the directed flow and its value becomes less then few percent 
of the angular momentum  accumulated by participants and even becomes negative when 
the antiflow occurs. The dominant part of the participant angular momentum is accumulated in 
the fragmentation regions at the freeze-out stage.


\begin{acknowledgments} 
Fruitful discussions with O.V. Teryaev and V.D. Toneev are gratefully acknowledged.
This work was carried out using computing resources of the federal collective usage center ``Complex for simulation and data processing for mega-science facilities'' at NRC "Kurchatov Institute", http://ckp.nrcki.ru/.
Y.B.I. was partially supported by the Russian Foundation for
Basic Research, Grants No. 18-02-40084 and No. 18-02-40085. 
A.A.S. was partially supported by  the Ministry of Education and Science of the Russian Federation within  
the Academic Excellence Project of 
the NRNU MEPhI under contract 
No. 02.A03.21.0005.  
\end{acknowledgments}

\end{document}